\documentclass{kluwer}
\usepackage{epsfig}

\begin{document}

\begin{article}
\begin{opening}

\title{New photometric models of galactic evolution applied  to the HDF}

\author{Gian Luigi \surname{Granato} \email{granato@pd.astro.it}}
\institute{Osservatorio Astronomico di Padova - Italy}
\author{Laura \surname{Silva} \email{silva@sissa.it}}
\author{Luigi \surname{Danese} \email{danese@sissa.it}}
\institute{SISSA - Trieste - Italy}
\author{Giulia \surname{Rodighiero} \email{grodig@pd.astro.it}}
\author{Alberto \surname{Franceschini} \email{franceschini@pd.astro.it}}
\institute{Dipartimento di Astronomia Univ.\ Padova - Italy}
\author{Gianni \surname{Fasano} \email{fasano@pd.astro.it}}
\author{Alessandro \surname{Bressan} \email{bressan@pd.astro.it}}
\institute{Osservatorio Astronomico di Padova - Italy}

\begin{abstract}

We summarize our modelling of galaxy photometric evolution  (GRASIL
code). By including the effects of dust grains and PAHs molecules in a two
phases clumpy medium, where clumps are associated to star forming regions, we
reproduce the observed UV to radio SEDs of galaxies with star formation rates
from zero to several hundreds $M_\odot \mbox{yr}^{-1}$.

GRASIL is a powerful tool to investigate the star formation, the initial mass
function and the supernovae rate in nearby starbursts and normal galaxies, as
well as to predict the evolution of luminosity functions of different types
of galaxies at wavelengths covering six decades. It may be interfaced with
any device providing the star formation and metallicity histories of a
galaxy.

As an application, we have investigated the properties of early--type
galaxies in the HDF, tracking the contribution of this population to the
cosmic star--formation history, which has a broad peak between z=1.5 and 4.
To explain the absence of objects at z $\gsim$ 1.3, we suggest a sequence of
dust--enshrouded merging--driven starbursts in the first few Gyrs of galaxies
lifetime.

We are at present working on a complementary sample of late type objects
selected in a similar way.

\end{abstract}

\end{opening}

\section{GRASIL}

Standard spectrophotometric synthesis consists  in summing the
spectra of all generations of stars (Simple Stellar Populations, SSPs) of
appropriate age and metallicity, as provided by the star formation history of
the galaxy. This simple procedure neglects the complexities (e.g.\ dependence
on geometry) introduced by a dust rich ISM, whose presence is probably the
rule for star forming systems.

To properly reproduce the spectral properties of galaxies from UV to sub--mm
we have included in our modelling the effects of dust in three environments:
(1) envelopes of AGB stars, (2) dense star forming molecular complexes
(MCs=HII regions+Giant Molecular Clouds), and (3) diffuse ISM (cirrus).

AGB dusty shells are directly included in the spectra of SSPs according to
the prescriptions given by Bressan, Granato \& Silva 1998. As for (2) and
(3), see the scheme sketched in Figure 1. The ISM of the axially simmetric
galaxy is divided in a dense phase associated with younger stellar
generations (MCs) and a diffuse phase. The association between dense dust and
young stars stems from the fact that we decrease with the age of each SSP the
fraction of its light radiated inside the cloud (Figure 2). The radiative
transfer is solved whenever necessary.

\begin{figure}[h]
\centerline{\epsfig{file=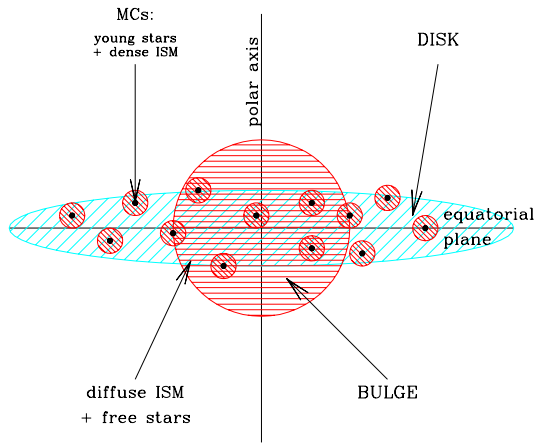,width=8cm}}
\caption{Scheme of the galaxy model, including both
a smooth medium and clumps (MCs) associated with star formation. The 
most general geometry consists in a superposition of a bulge and disk 
component.}
\end{figure}

\begin{figure}[h]
\centerline{\epsfig{file=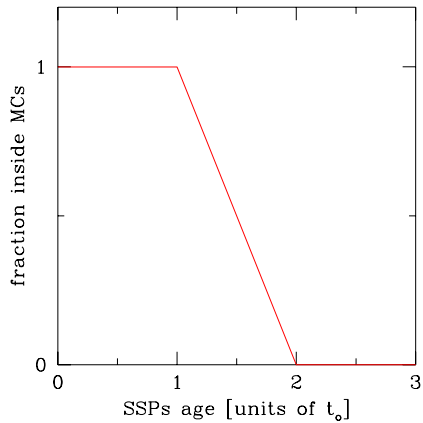,width=6cm}}
\caption{The association between 
dust clumps and young stars stems from the fact that we decrease with 
the age of each SSP the fraction of its light radiated inside the cloud.}
\end{figure}

\begin{figure}[h]
\centerline{\epsfig{file=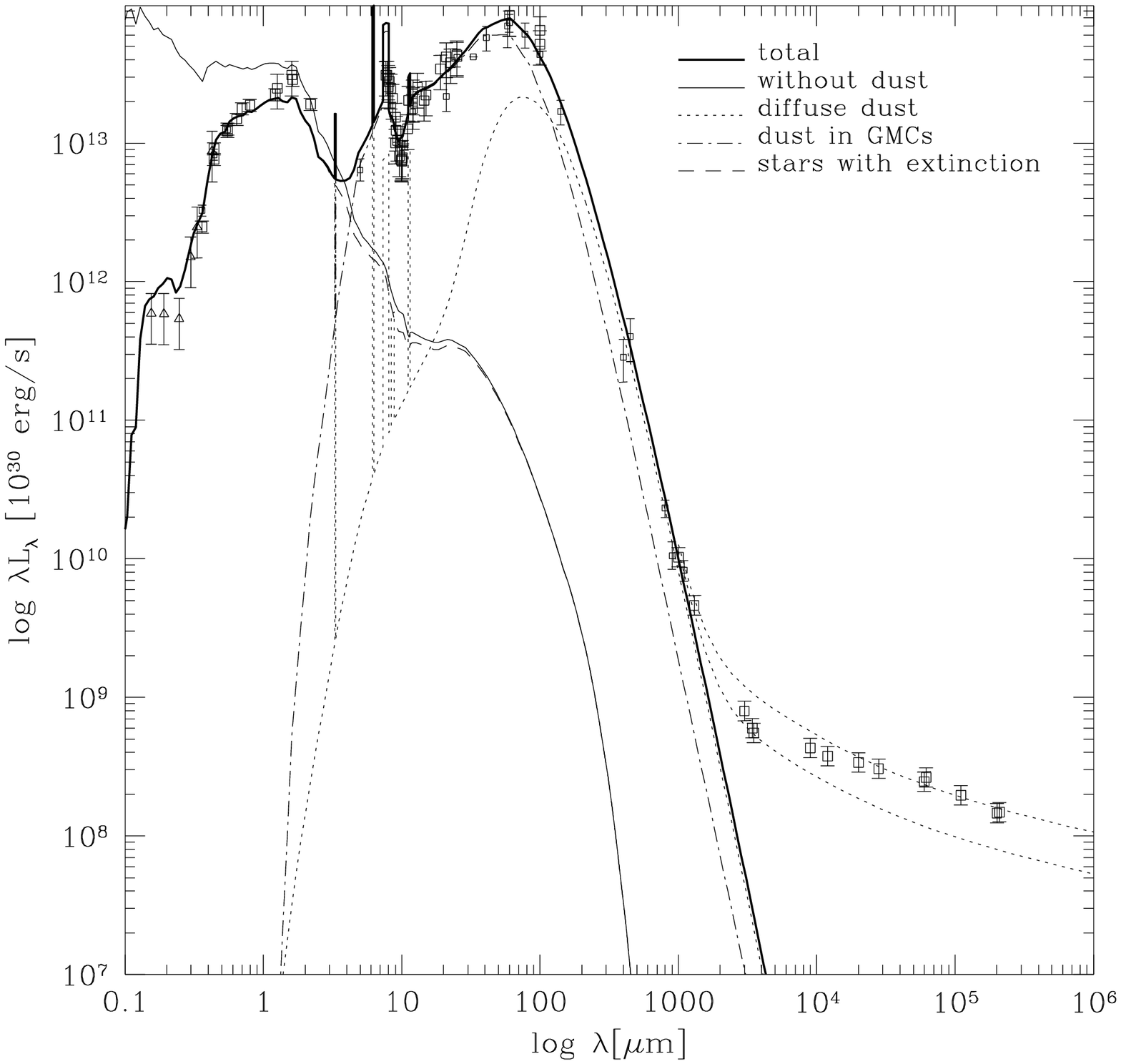,width=8cm}}
\caption{The observed radio to UV SED of the archetypal starburst galaxy M82
reproduced by GRASIL.}
\end{figure}

The resulting code (GRASIL) may be used as the observational interface of any
code providing the history of SF and eventually of metallicity of the galaxy
(see for instance Silva et al.\ 1999).

The observed UV to radio SEDs of different types of galaxies, from quiescent
ellipticals to massively star--froming ULIRGs, are remarkably well reproduced
with our mdel (see Figure 3 for an example and Silva et al.\ 1998 for
details).

\section{The contribution of field galaxies to the cosmic SF history.}

\begin{figure}[h]
\centerline{\epsfig{file=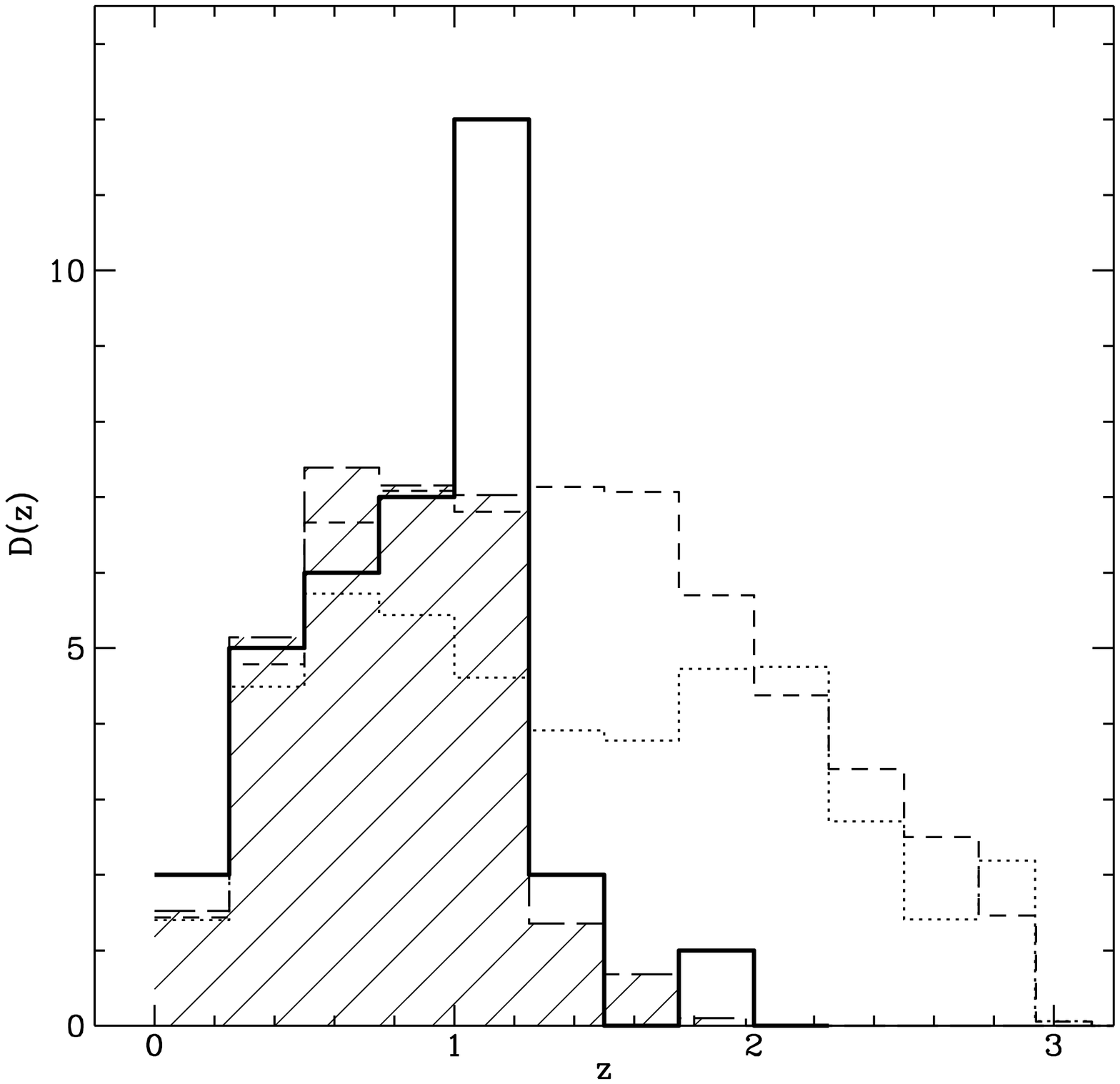,width=8cm}}
\caption{Continuous line: observed z--distribution of our sample 
of early type galaxies; dotted and shaded lines: model predictions 
without dust obscuration during the star--forming phase; shaded region:
model prediction including dust obscuration during star--formation.}
\end{figure}

In the past few years many efforts have been devoted in deriving the cosmic
history of star formation and/or metal production, which may constrain models
of structure formation. These derivations are mainly based on optically
selected high--z galaxies, in particular by means of the Lyman drop--out
technique. A well recognized problem is that optical surveys are strongly
biased against very old or very dusty galaxies.

We have thus pursued a complementary approach: a study of the stellar
populations of K band selected samples (one of early type and another of late
type galaxies) in the HDF, as rapresentative of corresponding field
populations. The K--band selection has two advantages: (1) it minimizes
the effects of K- and evolutionary corrections and (2) the flux at these
wavelenghts is contributed mainly by stars dominating the barionic mass.

\subsection{Early type galaxies}

The early type sample consists in 35 objects with K $\leq$ 20.15, z $\lsim$
1.3, and with the bulk of the light distribution well described by $r^{-1/4}$
profiles. The objects appear to be essentially free from dust obscuration. As
a consequence the 7 bands photometry (4 HST + 3 near--IR) from 0.3 to 2.2
micron allows to date the dominant stellar populations, which are found at z
$\lsim$1.3 to have a quite wide range of ages (typically from 1.5 to 3 Gyr).
Also the bright end of the E/S0 population appears to be mostly in place by
that cosmic epoch (see Franceschini et al.\ 1998 for details).

Having derived from the SED fitting procedure a good guess of the SF history
for each sample galaxy, we have extimated the contribution of the population
to the cosmic history of SF. This is compared with that derived from optical
surveys in Figure 6. It is clear that a significant fraction of light emitted
during the star--formation phases of early--type galaxies has been lost in
the optical, probably because obscured by dust. This possibility could
explain also the sharp cutoff at z $\sim$ 1.3 in the z distribution of our
sample galaxies (Figure 4).

The IR/sub--mm extragalactic background may be a trace of this
dust--extinguished SF, while a direct testing requires powerful dedicated
instruments.

\begin{figure}[h]
\centerline{\epsfig{file=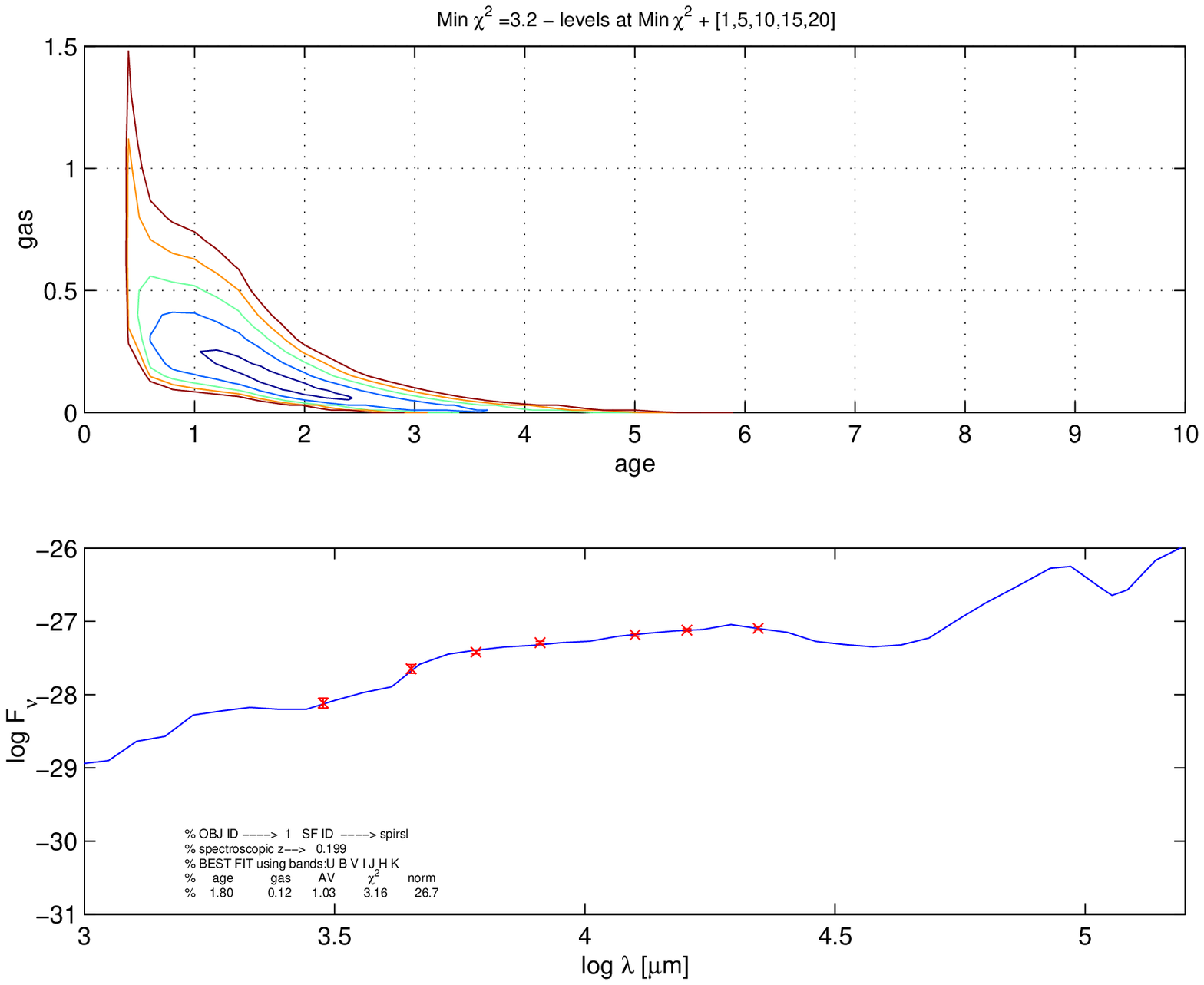,width=12cm}}
\caption{Best fit and $\chi^2$ contours for a typical object in
the late type sample: acceptable
fits have ages ranging from 0.5 to 3 Gyr, by adjusting the gas
content and thus the amount of extinction}
\end{figure}

\begin{figure}[h]
\centerline{\epsfig{file=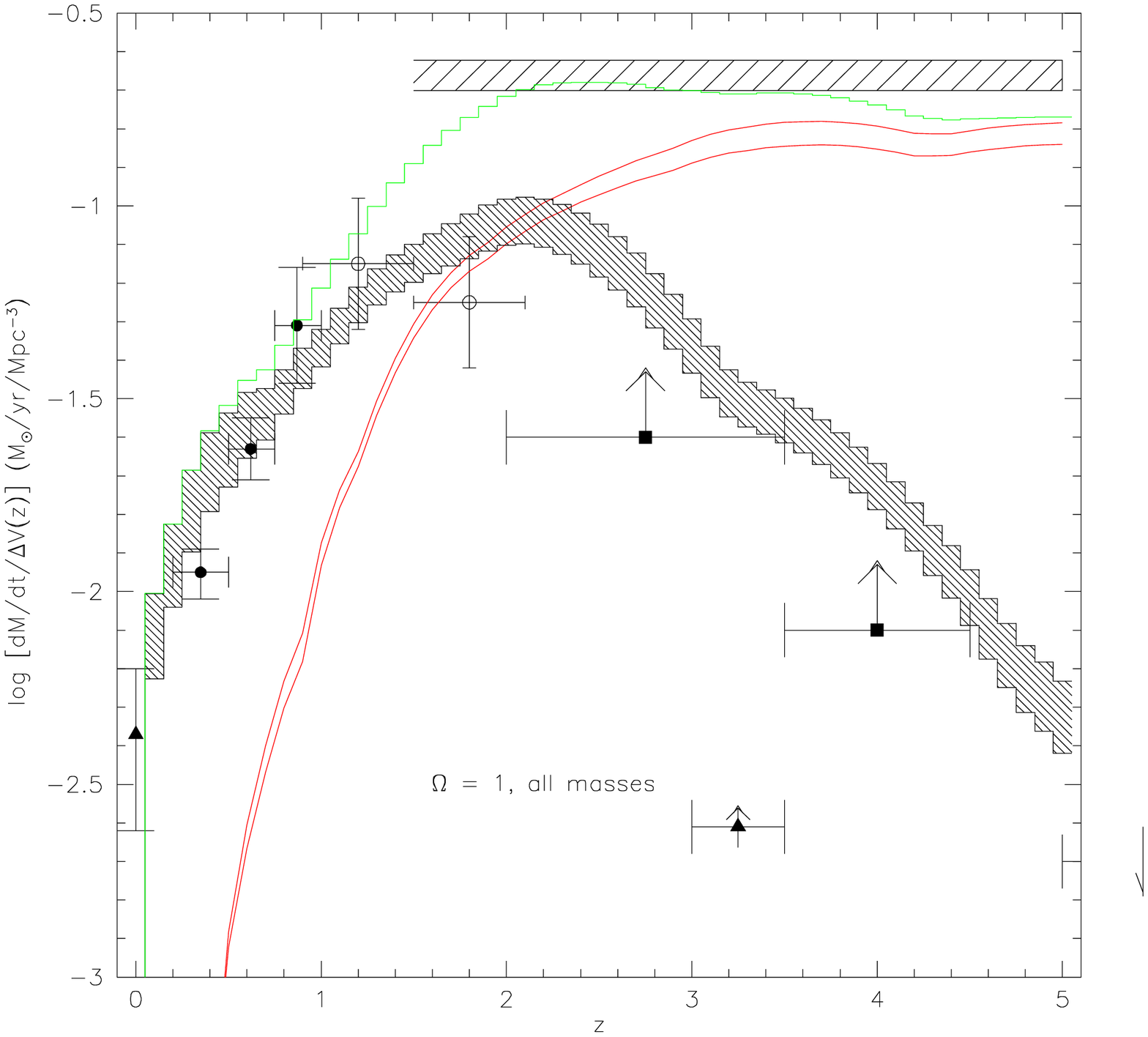,width=8cm}}

\caption{Contribution of field galaxies to the cosmic SF history:
the two lower solid line are upper and lower limit for field ellipticals,
the lower shaded area is bounded by two different estimates for field
late type, and the upper solid line is the sum of the previous two, which
turns out to be in good agreement with average determinations based on
metal abundance in clusters (horizontal shaded area).}

\end{figure}

\subsection{Late type galaxies}

A complementary effort we are carrying on, is devoted to a sample of
52 late type (spiral and irregular) galaxies with $K\leq 20.5$. In this case
the interpretation is more complex because the light emitted by these objects
is clearly dust extincted to some degree. As an example we show in Figure 5
the best fit we obtain, for a typical object in the sample, together with its
$\chi^2$ contours. The latter are intended to give a feeling of the level of
degeneracy between age and gas content: even assuming a specific model galaxy
with a fixed star formation history typical for spirals, it is possible to
obtain acceptable fits in a range of ages from 0.5 to 3 Gyr at least, simply
adjusting the gas content and thus the amount of extinction. Only a good
coverage of the IR region (in particular above $\sim 40 \mu$m), where  dust
radiates the absorbed energy, would allow to discriminate between the many
viable possibilities on the basis of optical and near--IR data alone.
Therefore our preliminary determination of the contribution of these objects
to the SF history of the universe, also shown in Figure 6, is prone
to larger model uncertainties, presently under study.

\end{article}

\end{document}